\documentclass[letterpaper]{article}
\usepackage{aaai}
\usepackage{times}
\usepackage{helvet}
\usepackage{courier}
\usepackage{graphicx}

\newcommand {\citeapos}[1]{\citeauthor{#1} (\citeyear{#1})}

\begin{document}
%
\title{An analysis of replies to Trump's tweets}
\author{Zijian An, Kenneth Joseph\\
Computer Science and Engineering, University at Buffalo\\
Buffalo NY, 14260\\
\{zijian,kjoseph\}@buffalo.edu
\\
}
\maketitle
\begin{abstract}
\begin{quote}
Donald Trump has tweeted thousands of times during his presidency.  These public statements are an increasingly important way through which Trump communicates his political and personal views. A better understanding of the way the American public consumes and responds to these tweets is therefore critical. In the present work, we address both consumption of and response to Trump's tweets by studying replies to them on Twitter.  With respect to response, we find that a small number of older, white, left-leaning, and female Americans are responsible for the vast majority of replies to Trump's tweets.  These individuals also attend to a broader range of Trump's tweets than the rest of the individuals we study. With respect to consumption, we note that Trump's tweets are often viewed not in isolation, but rather in the context of a set of algorithmically-curated replies. These replies may therefore color the way Americans consume Trump's tweets. To this end, we find some evidence that Twitter accounts see replies in line with their political leanings. However, we show that this can be largely, although not entirely, attributed to the fact that Twitter is more likely to show replies by accounts a user follows. As a basis for comparison, all results for Trump are compared and contrasted with replies to Joe Biden's tweets.
\end{quote}
\end{abstract}

\section{Introduction}

Since taking office on January 20, 2017, President Donald Trump has sent thousands of tweets.  The content of these tweets has varied widely, from mundane chatter to personal insults, from policy to misinformation \cite{joseph_polarized_2019}.  Taken collectively, however, their influence cannot be ignored.  For example, the volume and content of President Trump's tweets are associated with the performance of major stock indices \cite{stewart_volfefe_2019}.
And his insults towards the media and members of it are associated with a rise in harassment of and threats towards journalists \cite{follman_trumps_2018}.

\begin{figure}[t]
    \centering
    \includegraphics[width=.95\linewidth]{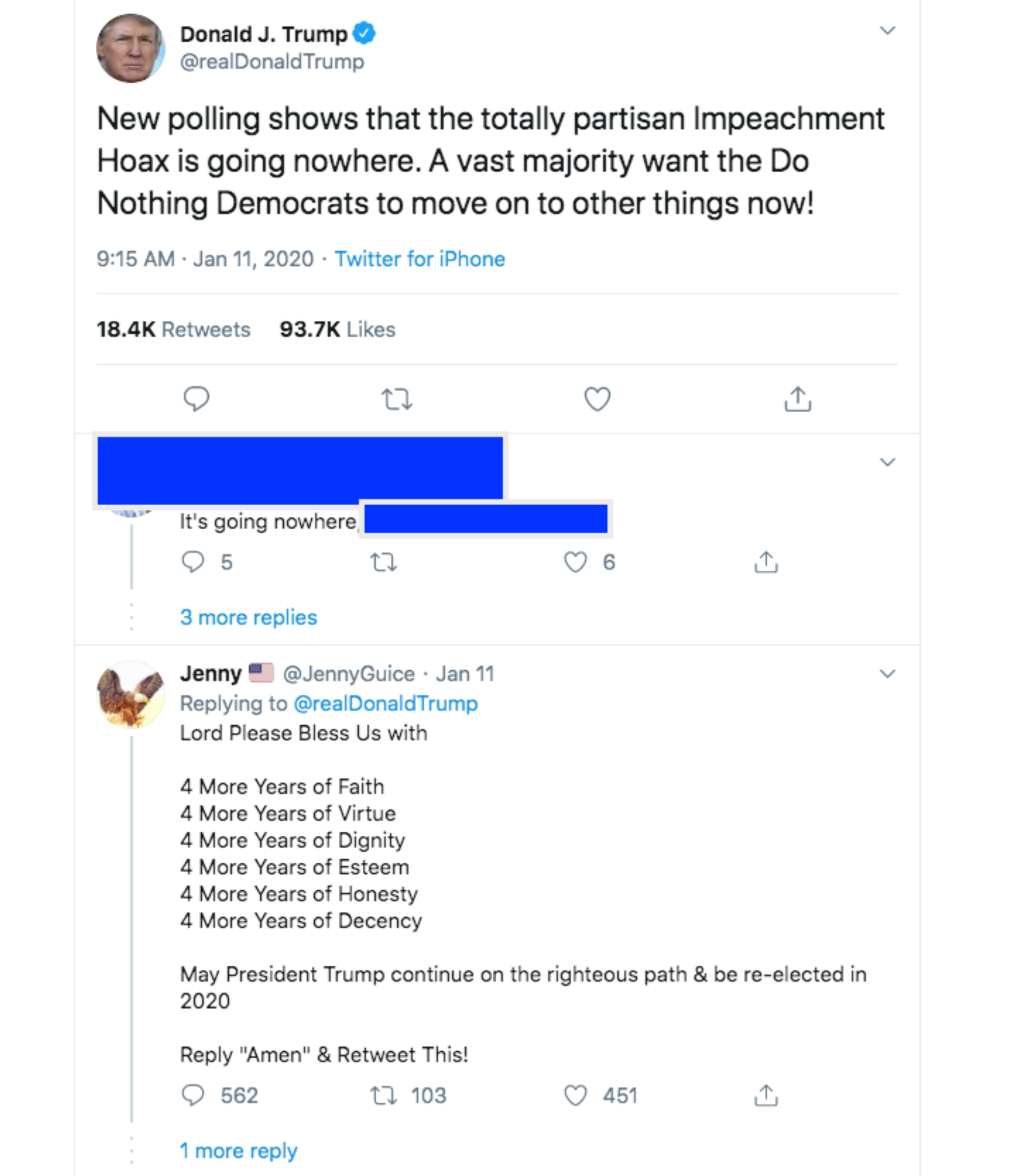}
    \caption{An example of the screen seen when one clicks on a tweet. Our first research questions asks, when do real Americans reply to Trump's tweets? Our second asks, when real Americans click on Trump's tweets, what do they see? Note that accounts with fewer than 10,000 followers are anonymized.}
    \label{fig:intro}
\end{figure}

Perhaps because of their incendiary content, President Trump's tweets typically generate a strong response from both supporters and detractors. One place where these responses are reflected is in the replies to his tweets.   Like Twitter in general \cite{ferrara_rise_2016}, and perhaps in particular with political content on the site \cite{bessi2016social}, at least some of the replies to Trump's tweets are from bots.  However, the space for replies underneath Trump's tweets, shown in an interface similar to that in Figure~\ref{fig:intro}, is also prime real estate for real people.  These individuals may want to express support, or to correct misinformed statements \cite{lafrance_first_2016}. Regardless of intent, these replies are an opportunity for individual Americans to express their political voice, and moreover, for it to be (potentially) heard by others.

However, we know little about who expresses their political voice in replies to President Trump's tweets, or for that matter, any other politician.  The first research question addressed in the present work is thus simply, \emph{who are the individuals that reply to Trump's tweets?}  To answer this, we use data from approximately 1.5M members of the American public who have a Twitter account and that can be linked to a voter registration record. Hereafter, we refer to this dataset of Twitter users linked to voter registration records as the \emph{panel} and a user within it as a \emph{panel member}.

With respect to this question, our findings are two-fold. First, replies to Trump's tweets by members of the American public are highly concentrated. Only 6.6\% of panel members have ever replied to one of Trump's tweets, and 80\% of all replies come from 0.7\% of the panel. Second, while the most active panel members reply to---and thus attend to---both popular and unpopular tweets from Trump, the vast majority of panel members reply only to his most popular tweets. Responses and attention to Trump by the majority of panel members are thus concentrated in a small subset of the content he has produced.  

This finding underscores a broader point that while most Americans are aware of the content of Trump's tweets, they are accessing this content indirectly. According to a 2018 Gallup poll \cite{newport_deconstructing_2018}, only 8\% of Americans follow Trump on Twitter.  In contrast, the same poll found that 53\% of Americans report seeing, reading or hearing ``a lot'' about Donald Trump's tweets, and another 23\% report they see, read, or hear a ``fair amount.'' The vast majority of Americans thus are consuming the content of President Trump's tweets, even if the tweets are not actively scrolling across their Twitter feed. 

How does one see Trump's tweets if not in their Twitter feed? One critical avenue is through cable television and/or online print media, which often promulgates Trump's social media musings \cite{pickard2017media}.  But, when an individual wishes to see one of Trump's tweet for themselves, they almost certainly do so by clicking on a link to that tweet.  In this case, they are exposed to the tweet through an interface that looks something like that in Figure~\ref{fig:intro}. 

As is clear, people who go to a specific tweet from President Trump see not only the content of that tweet, but also replies to it. Again, little is known about which of the thousands of replies sent to each Trump tweet are actually \emph{seen} when a tweet is viewed. Twitter's official documentation\footnote{https://help.twitter.com/en/using-twitter/twitter-conversations, accessed January 10th, 2020} states only that ``[r]eplies are grouped by sub-conversations'' using a ranking system, and that ``when ranking a reply higher, we consider factors such as if the original Tweet author has replied, or if a reply is from someone you follow.''   This documentation leaves open the possibility that organizational and bot-like accounts may be able to manipulate an algorithmic ranking system to appear more frequently than other replies, and that an individual with a particular partisan leaning may see replies aligned with their existing political views.  This would suggest the existence of an algorithmically induced ``filter bubble'' \cite{pariser_filter_2011}.

This reply interface therefore raises two important questions that are addressed in this work. First, are the voices of average Americans actually \emph{heard}, or are they drowned out by bots and organizational accounts? Second, when different people click on one of President Trump's tweets, are the replies they see shaped in a partisan way? That is, do Democrats always see Trump's tweets framed within a bevvy of negative responses, and Republicans, within a collection of accounts cheerleading his every move?

To address these questions, we carry out a simulation of six different Twitter accounts representing archetypes of heavily active and moderately activity Republican and Democrat-leaning panel members, as well as control accounts representing individuals who might use Twitter very sparingly. We simulate what these accounts would see when they click on several thousand of Trump's tweets. We then use this output to study who in our panel is heard, and whether individuals with different partisan leanings may be more likely to see replies that align with their political beliefs because of the ranking approach used by Twitter.

We present two main results from this portion of our analysis. First, the average American is not entirely drowned out---over one in five of Trump's tweets accessed by our simulated accounts display a response from at least one panel member. Second, we find that while simulated accounts do see replies that are biased along partisan lines, the effects are typically quite small, emerge mostly for replies to Trump, relative to Biden, and appear to be driven almost exclusively by the fact that users are more likely to see replies from other accounts they follow. 

Finally, it is important to ask about the generalizability of our findings to other politicians. Are our findings related to who replies and who sees which replies specific to Trump alone, or are they symptomatic of a broader pattern of behaviors towards American politicians? To address this question, we also analyze replies to tweets from Joseph Biden, Trump's competitor in the 2020 U.S. election. We find that replies to Trump are orders of magnitude more prevalent than replies to Biden, emphasizing the need to center Trump in our analyses.  Despite this difference in magnitude, however, we do find that nearly all main results for the research questions above hold for Biden as well. This suggests that patterns of who replies to Trump and who is seen when individuals click on Trump's tweets may generalize to other high-profile American politicians and responses to them on Twitter.\footnote{Code, and certain subsets of data, necessary to replicate and/or extend our work are available at \textit{https://github.com/kennyjoseph/icwsm\_trump\_replies}.}


\section{Related Work}
Our efforts relate to prior work on President Trump's tweets, the broader study of political content on Twitter, and research on online political participation.

\subsection{On President Trump's Tweets}
Quantitative analysis of Trump's tweets has suggested that he frequently expresses content relating to policy, but also frequently leverages Twitter for personal attacks and attacks on the media \cite{anderson_tweeter--chief_2017}. Such work has largely been carried out in non-academic outlets.  For example, journalists at The Washington Post and The New York Times have respectively catalogued misinformation in Trump's tweets \cite{kessler_trump_2019} and instances of personal attacks \cite{lee_487_2016}. The most relevant academic work is that of \citeapos{joseph_polarized_2019}, who calculate measures of support for Trump's tweets based on reply and retweet counts. Like us, these authors found that patterns relevant to Trump's tweets were also reflected, although to a less extreme extent, with other American politicians. However, this prior work does not investigate patterns in who replies, instead assuming that any reply is simply a negative indicator of support. Further, it does not consider which replies are seen.

In addition to this quantitative work, which is more aligned with the present efforts, qualitative scholars have also begun to study President Trump's tweeting behavior. For example, \citeapos{ott_age_2017} uses a theoretical analysis of Trump's use of Twitter to suggest ways in which Twitter has polluted socio-political discourse. And \citeapos{ross_discursive_2018} use qualitative methods to emphasize the ways in which President Trump strategically uses the concept of fake news to promulgate dis- and misinformation, and to encourage additional loyalty from his supporters.  The present work compliments these deeper analyses of specific tweets with a large-scale quantitative approach.

\subsection{On Politics and Twitter}

A significant literature exists showing that Twitter has been widely used as a means to express political views \cite{lin_rising_2014}. 
Most relevant to our efforts are two subsets of this prior work.
The first subset, relevant to our first research question, emphasizes inequality in the distribution of political voice and attention on Twitter.   During the 2015 British general election, a highly active subset of users, typically supporting nationalist parties, controlled much of the discussion \cite{segesten_typology_2017}. During the 2012 U.S. presidential debates, \citeauthor{lin_rising_2014} (\citeyear{lin_rising_2014}) found that the top 25\% people account for approximately 75\% of all retweet activity. And \citeauthor{grinberg_fake_2019} (\citeyear{grinberg_fake_2019}) find that during the 2016 presidential election, 0.1\% of the users they studied accounted for around 30\% of all shares of URLs containing political content, and 80\% of all political fake news shared. The inequality we find in responses to Trump's tweets by average Americans align with these prior observations in other contexts.

The second subset, relevant to our analysis of which replies are seen, pertains to the existence (or lack thereof) of what \citeapos{pariser_filter_2011} calls filter bubbles. The notion of a \emph{political} filter bubble, or echo chamber \cite{barbera2015tweeting}, is that individuals who are right- (left-)leaning will be encircled only by other right- (left-)leaning individuals, sharing only right- (left-)leaning content. 
Given the algorithmic ranking of replies when a link to a tweet is clicked, the specific concern here is on the existence of \emph{algorithmically curated} political filter bubbles, where algorithms enforce and encourage interactions only with others who share the same political views.  

Empirical evidence of algorithmically-curated political filter bubbles is, at best, mixed.  \citeapos{kulshrestha_quantifying_2017} find, in line with assumptions of the filter bubble, that in Twitter's search interface, searches for left- (right-)leaning candidates lead to algorithmically-curated rankings that are more left- (right-)leaning. However, they also observe that this can be explained in large part by the fact that tweets from left- (right-)leaning users about left- (right-)leaning politicians tend to be more popular.  \citeapos{robertson_auditing_2018} find almost no evidence of algorithmically curated political filter bubbles in Google search, instead finding that most of the partisanship in search results can be explained simply by the partisanship of the query itself.

In each of these prior works, analysis of algorithmically curated filter bubbles relies on the ability to score content (a tweet, or a search result) or sources (a user sharing a tweet) on a left-to-right scale. Here, we adopt straightforward and well-established methods to score replies for partisanship according to both source and content. Our approach to scoring sources is taken directly from the work of \citeapos{barbera_less_2016}; our content-scoring approach is a straightforward application of modern NLP methods for text classification.

\subsection{Political Participation}
Finally, prior work on political participation provides insights into who the more active responders to Trump's (and Biden's) tweets may be, what their motivations might be, and what form of political participation these replies should be considered to be. A number of articles have shown that online political behavior is linked to other, more traditional forms of offline political participation \cite{ferrucci_civic_2019,skoric_social_2016}; thus, more active repliers are likely more politically active in general. They are also more likely to consume political content that reinforces their beliefs, relative to more neutral or opposing content, and to be more politically extreme \cite{feezell_predicting_2016}.

With respect to motivations, individuals who reply to Trump or Biden's tweets are likely stimulated to do so by factors that are both intrinsic (e.g. for personal satisfaction) and extrinsic (e.g. a desire for social approval) \cite{lilleker_what_2017}. However, extrinsic factors--- including a desire to disseminate information, present the self favorably, and to persuade others---are more likely to motivate individuals to participate in politics online \cite{lilleker_what_2017,winter_examining_2016}.  This drive to participate online may also increase the individual's offline social capital. \citeapos{gil_de_zuniga_social_2017} find that online social capital facilitates offline political participation, which is itself associated with offline social capital. Thus, it is possible that becoming influential online can lead to increased political status offline as well.  It is therefore important to understand potential influence of accounts engaging with Trump's tweets.

This relationship between online and offline social capital mirrors a close relationship between online and offline political participation \cite{gibson_conceptualizing_2013}. Indeed, communication scholars have increasingly argued that online political speech acts are an important form of political participation in and of themselves \cite{gil_de_zuniga_social_2017}.  Replying to politician's tweets falls under what \citeapos{gibson_conceptualizing_2013} calls the targeted and expressive modes of political participation. Targeted participation occurs when individuals interact directly with politicians, e.g., by calling their representatives or tweeting at them online. Expressive participation occurs when individuals express political sentiments more broadly, e.g. through yard signs or undirected tweets.


\section{Who Replies to Trump's Tweets?}

In this section, we provide an analysis of who replies to President Trump's tweets, showing that our main results also extent to another high-profile American politician, Senator Joe Biden. We first introduce the data used, and then provide an analysis of results. 

\subsection{Data}
We study replies to 11,381 of Trump's tweets from January 1st, 2018 through July 30th, 2020. As a point of comparison, we also study replies sent to 2,597 tweets sent by Joe Biden during the same time period. Replies were extracted from our panel of approximately 1.5M Twitter users that have been linked to voter registration records, who have sent at least 10 tweets, and who were active\footnote{That is, the account was not suspended, deleted, or protected} as of August of 2019. 

Voter registration records are used to determine the party registration, self-identified race/ethnicity, self-identified gender, and age of panel members. Tweets were collected from panel members on a biweekly basis, beginning in July of 2017, using the Twitter Search API. In total, we analyze 2.1M replies to Trump and 83,000 replies to Biden. Replies to Trump were sent by 90,279 panel members, replies to Biden were sent by 19,597 panel members. Only tweets of Trump and Biden's that were replied to by at least one panel member were analyzed. 

The method we use to link Twitter accounts to voter registration records is similar to those presented in prior work  \cite{barbera_less_2016,grinberg_fake_2019}. We start with approximately 400M Twitter accounts that shared a public tweet captured by the Decahose from 2014-2016, and a collection of approximately 300M voter registration records that we obtained from TargetSmart. In all cases, the voter registration records provide an individual's name and state. As noted, in many but not all cases, they also provide demographic information, including age, self-identified gender, self-identified race/ethnicity, and whether or not an individual is registered with a political party. 

We link a Twitter account to a voter record if both have the same listed name and location, if that name/location combination is unique, and if no one else on Twitter exists with the same name but an unidentified location.  Names and locations are matched only when exact links can be found. Notably, our approach relies on finding individuals who express their full name and location on Twitter, and whose names are unique within their state. This biases our sample towards people who provide their real names and locations, and thus likely against observing more virulent content. However, the sample we use is still representative, in terms of typical demographic variables, of the broader American public on Twitter as represented in survey data from PEW \cite{greenwood_social_2016}. Full details on the methodology used to link voter registration records to Twitter accounts, including an extensive evaluation showing that the approach is likely over 90\% accurate, is available in prior work \cite{grinberg_fake_2019}.  Note that we link approximately 0.5\% of the available voter records in our voter record dataset. Using statistics about the percentage of Americans who use Twitter, and the percentage of Americans registered to vote, we estimate that our panel contains roughly 1-3\% of all voting Americans on Twitter.

Our approach has been approved by the IRB at Northeastern University, and does not include data from any individual who has in any way (e.g. by changing a letter of their name) sought to conceal their full name or location on Twitter.  Because of this, our approach also does not violate Twitter's Terms of Service. Still, we acknowledge that there are important ethical questions surrounding how and when this data should be used. In the context of the present work, we believe that the minimal risks posed to individuals in our panel are acceptable, given the importance of understanding potential discrepancies and patterns in how their voices are expressed and heard online.

\subsection{Results}

\begin{figure}[t]
    \centering
    \includegraphics[width=.47\textwidth]{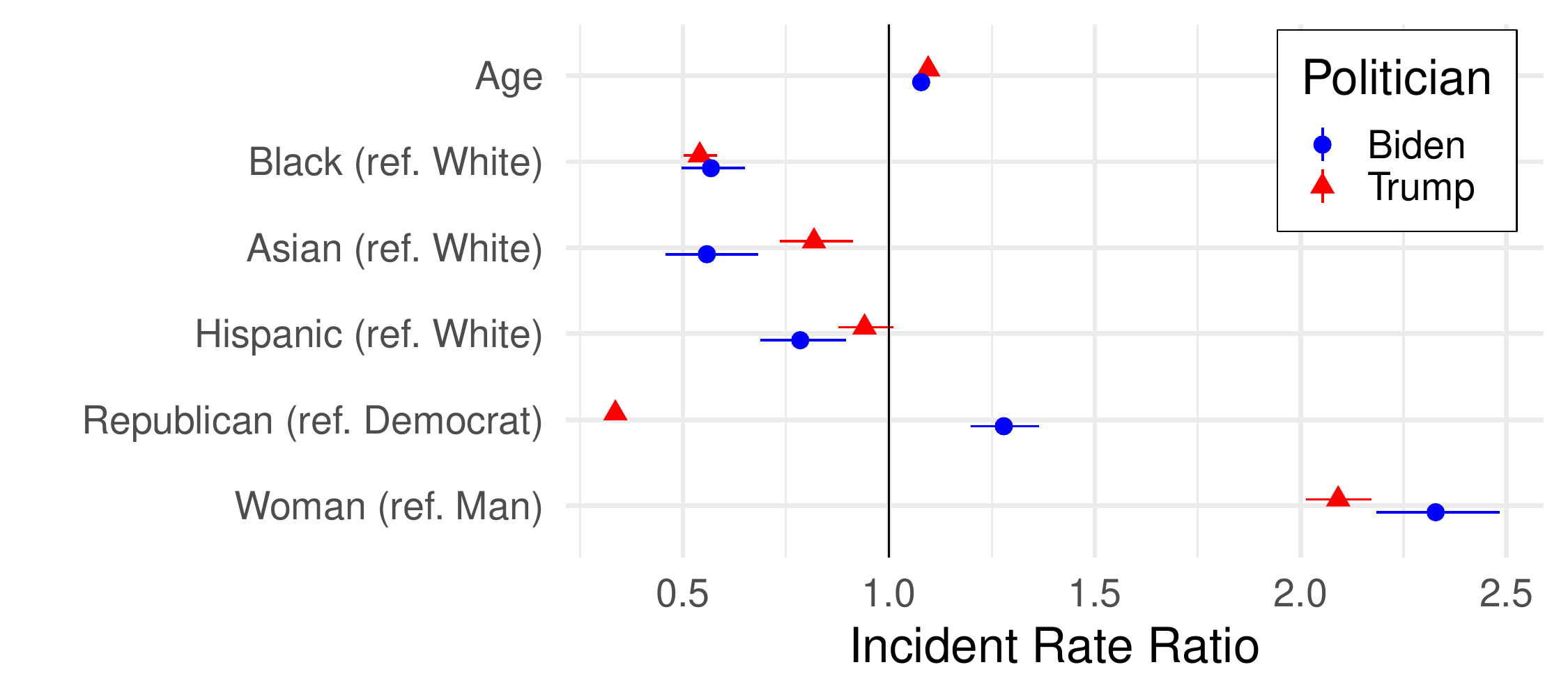}
    \caption{Results of a negative binomial regression predicting the number of replies an individual sends to any of Trump's tweets (red coefficients) or Biden's tweets (blue coefficients). The x-axis depicts incidence rate ratios for each coefficient. Not shown is the control variable for number of total tweets sent by an individual.}
    \label{fig:reg_results}
\end{figure}

As shown in Figure~\ref{fig:reg_results}, individuals who reply to both Trump's tweets and Biden's tweets more frequently are more likely to be female, white, and older. While demographic differences in political participation are notoriously difficult to measure \cite{anoll2018makes} and can vary across issues \cite{holbrook_racial_2016}, these demographics constitute individuals who are traditionally expected to be more politically active \cite{pew_research_center_10._2018}.  The only salient difference between the demographics of individuals replying to Trump as opposed to Biden was that, as expected given previous work \cite{garimella_quote_2016,joseph_polarized_2019}, Republicans were more likely to reply to Biden, and Democrats more likely to reply to Trump. 

Results were obtained using a generalized linear model with a negative binomial link function, where the dependent variable is how many times a user replies to Trump or Biden's tweets. Coefficients for independent variables are displayed as incident rate ratios. Incident rate ratios provide the multiplicative effect of the covariate over the reference condition. For example, on average, a women in our data replies to Trump around 2.1 times for each reply from a man, controlling for the other factors in the model. Independent variables include a control for the number of total tweets from the individual, as well as predictors for age, race/ethnicity, political party, and gender.\footnote{Note that for the results in Figure~\ref{fig:reg_results} only, we considered only the 31\% of individuals in our dataset for whom we have complete demographic information. All other results in this paper include all panel members.}

\begin{figure}[t]
    \centering
    \includegraphics[width=.48\textwidth]{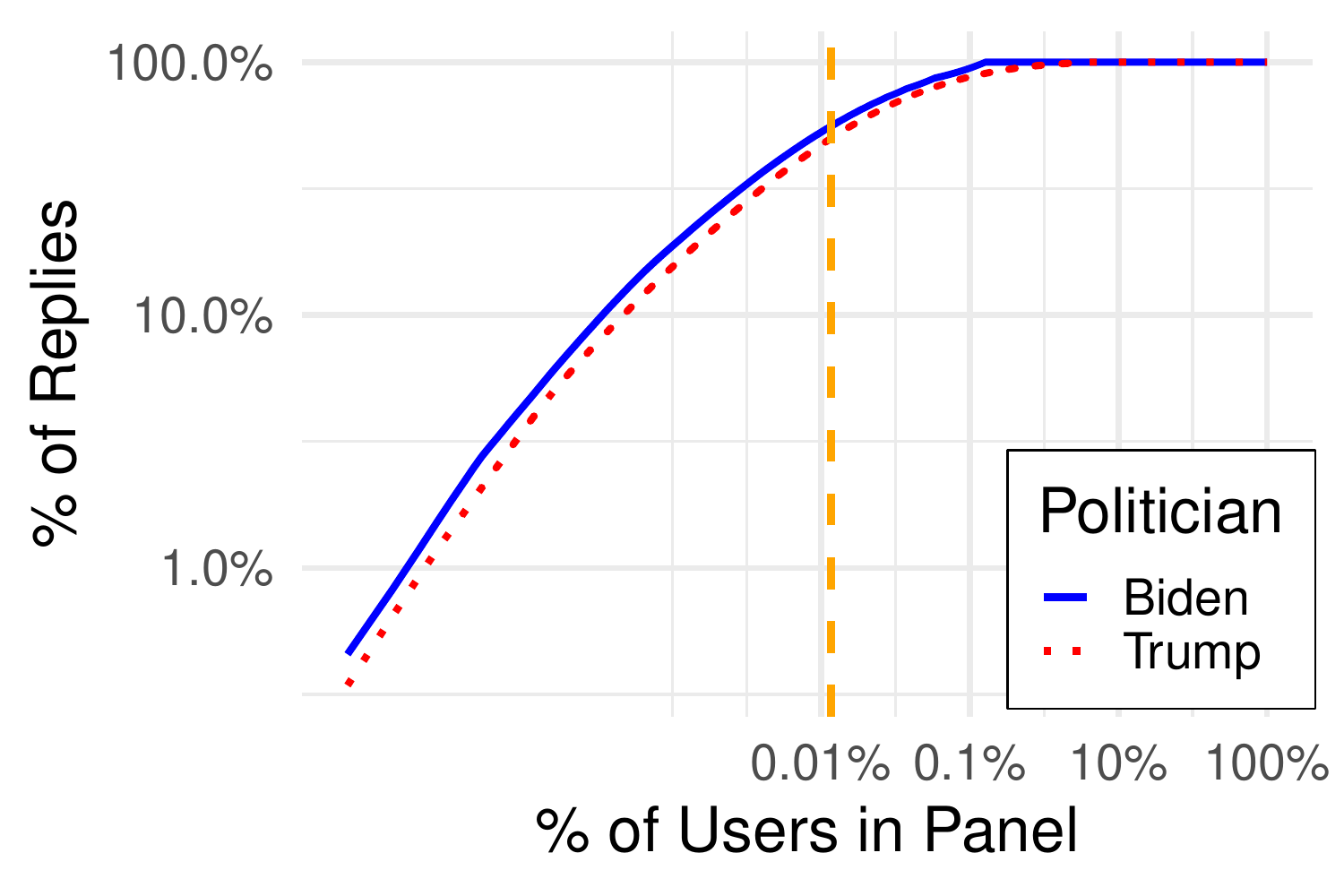}
    \caption{ Empirical cumulative distribution function displaying the percentage of users in our panel that accounted for a given percentage of replies to Trump (red line) or Biden (blue line). The orange, dotted vertical line indicates the point at which the given percentage of users (0.12\%) accounts for 50\% of all replies to Trump, similar to the point for replies to Biden (0.09\%) }
    \label{fig:ccdf}
\end{figure}

As shown in Figure~\ref{fig:ccdf}, however, the vast majority of replies were concentrated within a few individuals.  Only 6.6\% of the users in our dataset ever replied to Trump, and of those, a mere  1.9\% of all individuals who replied at least once accounted for over half of all replies. These metrics are similar for Biden; only 1.3\% of users ever replied to a Biden tweet, and of those, only 6.7\% accounted for over half of all replies.   Political voice in the panel, as expressed in replies to both Trump's tweets and Biden's tweets, was therefore highly concentrated within a small group of panel members. 

To further explore differences between highly active versus less-active panel members, we split them into \emph{Active} and \emph{Non-active} repliers to Trump and Biden's tweets.  Active repliers are the 1,794 (1,320 for Biden) highly engaged panel members that account for 50\% of all replies to Trump's tweets, non-active repliers are the other 88,485 (18,277) individuals who replied to at least one of Trump's (Biden's) tweets. Splitting the data at this 50\% mark enables us to study how these two types of individuals distribute their replies to Trump's tweets. In particular, we would like to know the distribution of attention across all of Trump's tweets for each group. 

Replies from Non-active users were significantly more concentrated in Trump's most popular tweets than replies from active users.  The Gini coefficient, which measures the concentration of a variable, of the reply distribution across Trump's tweets was 0.71 (95\% bootstrapped confidence interval [0.705 0.713]) for Non-Active users and .62 [.617,.623] for active users. This inequality means that the majority of Trump's tweets had a majority of replies from Active users, even though, by construction, Active and Non-active users accounted for the same number of replies. Specifically, an average of 54.6\% of replies to a given tweet were by active accounts, and 65.1\% of Trump's tweets had more replies from active users than non-active users.
These patterns are also observed in data on replies to Biden's tweets. The Gini coefficient of the reply distribution across Biden's tweets was .76 [.750,.767] for Non-active users and .70 [.693, .707] for active users, in turn resulting in 54.5\% of replies to an average Biden tweet coming from active users, and 55.1\% of his tweets seeing more replies from active than Non-active users.

Political attention for a select few highly active individuals thus was spread more broadly across Trump and Biden's tweets. In contrast, attention, as quantified by replies, from the vast majority of individuals in our panel was focused only on the most popular tweets. These popular tweets, in turn, tend to be the most incendiary \cite{joseph_polarized_2019}. For example, around 1 out of 300  replies from Non-active accounts were directed at a single tweet in which Trump, with no evidence, suggested a video of a Black Lives Matter protester in Buffalo, NY being shoved by police was a ``set-up''.\footnote{https://twitter.com/realDonaldTrump/status/1270333484528214018}. In contrast, only 1 in 1,000 replies from active accounts were directed at this tweet. 


\section{Who Sees Which Replies to Trump's Tweets?}

We now turn to questions about which replies to Trump's (and Biden's) tweets are seen. Ideally, we would do so by gathering data on the replies seen by real Twitter users (e.g. panel members). However, while scholars have developed various approaches to using Twitter's API to estimate exposure of real people to Twitter feeds (e.g. \cite{grinberg_fake_2019}), such approaches do not extend to the viewing of replies.  Consequently, we chose to use an approach common in algorithmic auditing studies (e.g. \cite{robertson_auditing_2018}) where we create \emph{simulated} accounts that \emph{emulate} replies seen by real people.  Our simulation approach is limited in the sense that we can only simulate a small number of accounts, none of which are exactly representative of any single ``real person.'' However, by carefully constructing these accounts to be generally representative of particular types of people, our simulated accounts allow us to make narrow but well-supported statements about what certain kinds of real people might see.

\subsection{Data}

We create six different simulated Twitter accounts to emulate the replies that real people might see when they click on one of Trump or Biden's tweets.  These simulated accounts differ \emph{only in the accounts they follow}.  We hypothesized that the accounts followed by each simulated account might impact replies the account ``sees'' in two ways. First, as noted, Twitter curates replies shown to a particular user in part by privileging replies sent by other accounts that user follows.  Second, accounts followed by a simulated account may be used as implicit signals to determine which content to show even when there are no (or few) replies from the followed accounts. 

Two of the six simulated accounts are baselines, used to characterize individuals who use Twitter sparingly and whose accounts therefore have limited partisan signal. The first baseline account follows no other accounts, the second follows exactly one other account, \@realDonaldTrump (President Trump's most frequently used account). The final four accounts are created to be representative of users studied in the previous section who have replied to President Trump at least once: Active Democrats, Non-active Democrats, Active Republicans, and Non-active Republicans. Our interest in these accounts stems from the fact that in replying to Trump's tweets, these individuals are almost certain to have accessed the screen in Figure~\ref{fig:intro}.  For these accounts, we can therefore be more certain that they are exposed to the interface that we study.

We create one simulated account to represent a typical user for each of these four classes. To do so, we first collected all accounts followed by anyone in our panel who replied to at least one of Trump's tweets. Data was collected in December of 2018. We then identified the median number of accounts followed by users in each group. Finally, we then had the simulated typical account follow that many of the most heavily followed accounts by users in that group. For example, the median Active Democrat followed 638 accounts. Our Active Democrat simulated account therefore followed the 638 most commonly followed users by the Active Democrats identified in the previous section.  The median Active Republican, Non-active Democrat, and Non-active Republican followed 425, 427, and 388 other accounts, respectively.

Having set up the following relationships of the simulated accounts, we then used Selenium\footnote{https://selenium-python.readthedocs.io/} to simulate a real person clicking on tweets from both Trump and Biden. For each simulated account and each tweet, we recorded the replies that appear in the HTML. 

For each of the six simulated accounts, we collect six different \emph{samples} of reply tweets to reflect different \emph{time periods} at which an individual might view a politician's tweet. We first gathered five \emph{Real-time} samples. These Real-time samples gathered replies ``seen'' by the simulated accounts at exactly 10, 20, 30, 40, and 50 minutes after a tweet was sent by Trump (n=120) or Biden (n=80) from May through June of 2020. A final \emph{Complete} sample was collected in August of 2020 for replies seen for all of Trump's (n=9713) and Biden's (n=2556) tweets sent between 2018 and July of 2020 that received at least one reply from a panel member.  

For each tweet in each Real-time and Complete sample, we collect replies seen by each simulated account at approximately the same exact time. We scrape up to the top 10 replies. Finally, we remove from the sample any tweet from a user who was deleted or suspended in early August of 2020.  This accounted for approximately 3\% of the Complete sample, and approximately 15\% of the five Real-time samples. Below, we refer to any account shown to at least one of the simulated accounts within at least one sample as a \emph{replier}. In total, there are 99,156 such replier accounts.

\subsection{Methods}

Our analysis focuses on two main questions. First, do replies from panel members, studied above, actually appear in the replies seen by simulated accounts? Second, is there evidence of a partisan tilt in the replies shown to the simulated accounts? That is, are the accounts simulating viewing behaviors of (non-)active Democrats exposed to more left-leaning replies than (non-)active Republicans?

With respect to the latter question, prior work on the Twitter search page \cite{kulshrestha_quantifying_2017} has studied algorithmically curated filter bubbles at both the source level (in our case, \emph{who} the replier is) and the content level (\emph{what} they are saying in the reply).  We do the same, although of course for replies and not search results, and with slightly different methods.  Here, we first discuss how we measure source bias, and then how we measure content bias.  

\subsubsection{Quantifying Source Bias}

Our source bias metric is a measure of the average partisan leaning of all repliers (sources) observed for a given tweet, \emph{given only account-level information about the replier}.  More specifically, we compute the source bias for each politician's tweet, from a given sample, for a given simulated account, as \emph{the expected probability that any given replier observed is a Republican as opposed to a Democrat}.  

To do so, we first determine the likelihood that a given replier individually is a Republican using the approach of \citeapos{barbera_less_2016}. Similar approaches based on social connections have been widely adopted elsewhere in the literature \cite{wihbey2019social}, and the \citeapos{barbera_less_2016} approach is motivated by both empirical and theoretical underpinnings \cite{colleoni2014echo}.  To do so, we first build an L1-regularized logistic regression to predict the partisanship of the 502,956 panel members who are registered as either a Republican or Democrat. The independent variables used are the Twitter accounts the panel members followed. We then validate this model and apply it to the repliers.

More specifically, we first identity the 18,609 accounts followed by at least 1,000 panel members and at least 250 repliers. Doing so ensures that we focus on relatively popular accounts that are of shared use in predicting partisanship in both datasets. 
We then use cross-validation to tune the regularization parameter ($\lambda$) of the regression model on 90\% of the panel, and evaluate the model on a 10\% held-out test dataset.  Finally, we apply the model trained on the panel to the repliers.  

\begin{figure}
    \centering
    \includegraphics[width=.98\linewidth]{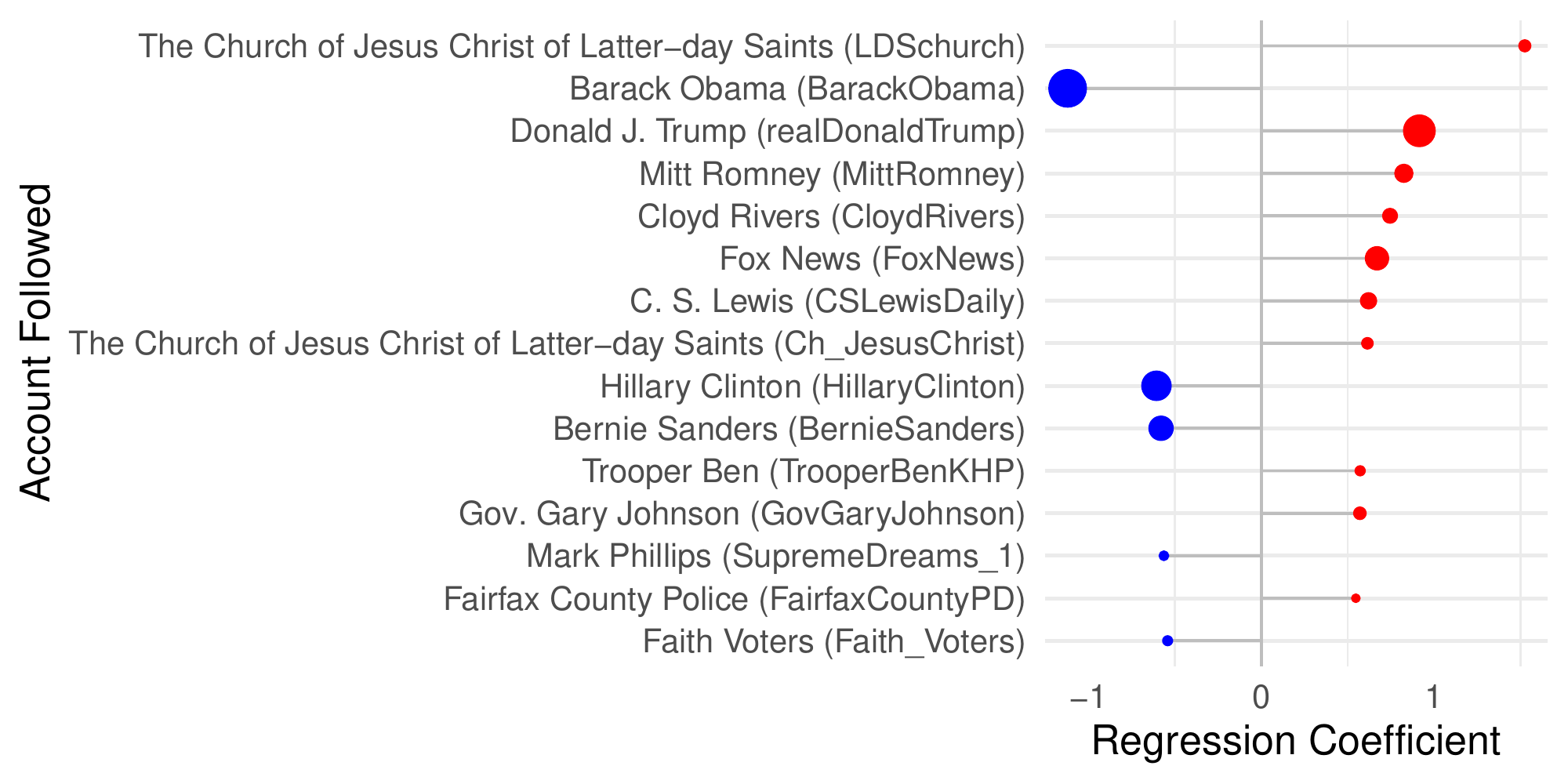}
    \caption{Top 20 coefficients, by absolute value, for an L1-regularized logistic regression to predict partisanship (registered Democrat or Republican) of Twitter users. Negative coefficients imply someone who follows this account is more likely to be a Democrat, positive, that they are more likely to be a Republican. The size of the point indicates the number of people following that account.}
    \label{fig:coef}
\end{figure}

With respect to model quality, we find that the model correctly predicted the registered party of 79.8\% of the users in our test set. This is a significant improvement over a baseline majority-only model\footnote{I.e. always predicting the majority class (Democrats)} performance of 63.3\%. This improvement over the baseline gives us confidence that the model can reasonably reflect partisan source bias. Coefficients from the model also provide face validity. Figure~\ref{fig:coef} presents the top 20 regression coefficients of the model in terms of absolute value, showing that the model prediction that a user who follows obviously left (right) leaning accounts will itself have a left (right) leaning partisan affiliation.

Given a measure of partisanship for each source, we then compute the source bias for a given tweet, seen by a given simulated account, for a given sample, by taking the average of the partisanship scores for each observed source.  This average gives us an (unbiased) estimate of the probability that any given replier for that tweet is a Republican.

\subsubsection{Quantifying Content Bias}

Our content bias metric is a measure of the average partisan leaning of the  all replies observed for a given tweet, \emph{given only the content of those tweets}.  More specifically, we compute the content bias for a given politician's tweet, from a given sample, for a given simulated account, as \emph{the expected probability that the text of the reply was produced by a Republican as opposed to a Democrat}.  

To do so, we first construct a balanced sample of the text of 158,018 replies sent by registered Democrats to Trump (16,759 to Biden) and 157,831 sent by registered Republicans to Trump (16,780 to Biden). We then train a machine learning model to predict the political leaning of the reply's sender based only on the content of the tweet. In doing so, we train the model to recognize content that is likely left- or right-leaning.  We train our model on 70\% of the data, and evaluate its performance using the remaining 30\%.  Note that we train separate models for replies to Trump and Biden.

The model we select is the transformer-based RoBERTa model \cite{liu2019roberta}, one of the most popular state-of-the-art natural language processing models for text classification. We use the \texttt{simpletransformers}\footnote{https://simpletransformers.ai/} package in python to fine-tune the base pretrained RoBERTa model for our task. Models are trained on a single NVIDIA Titan V GPU for 50 and 20 epochs and took approximately 1.5 and 6 hours to train for the Trump and Biden specific models, respectively.

Our Trump-specific content bias model correctly predicts the political affiliation of a user sending a reply 71\% of the time.\footnote{Note that since the dataset is balanced, baseline performance is 50\%}. Our Biden model performs similarly well, correctly predicting the political affiliation of the user 75\% of the time.
To assess content bias for a given tweet from Trump, seen by a given simulated account, for a given sample, we first apply our content bias model to each reply. We then take the average of the content bias scores for each observed reply. We perform the analogous step for replies sent to Biden's tweets.


\subsection{Results}

\begin{figure*}[t]
    \centering
    \includegraphics[width=.97\textwidth]{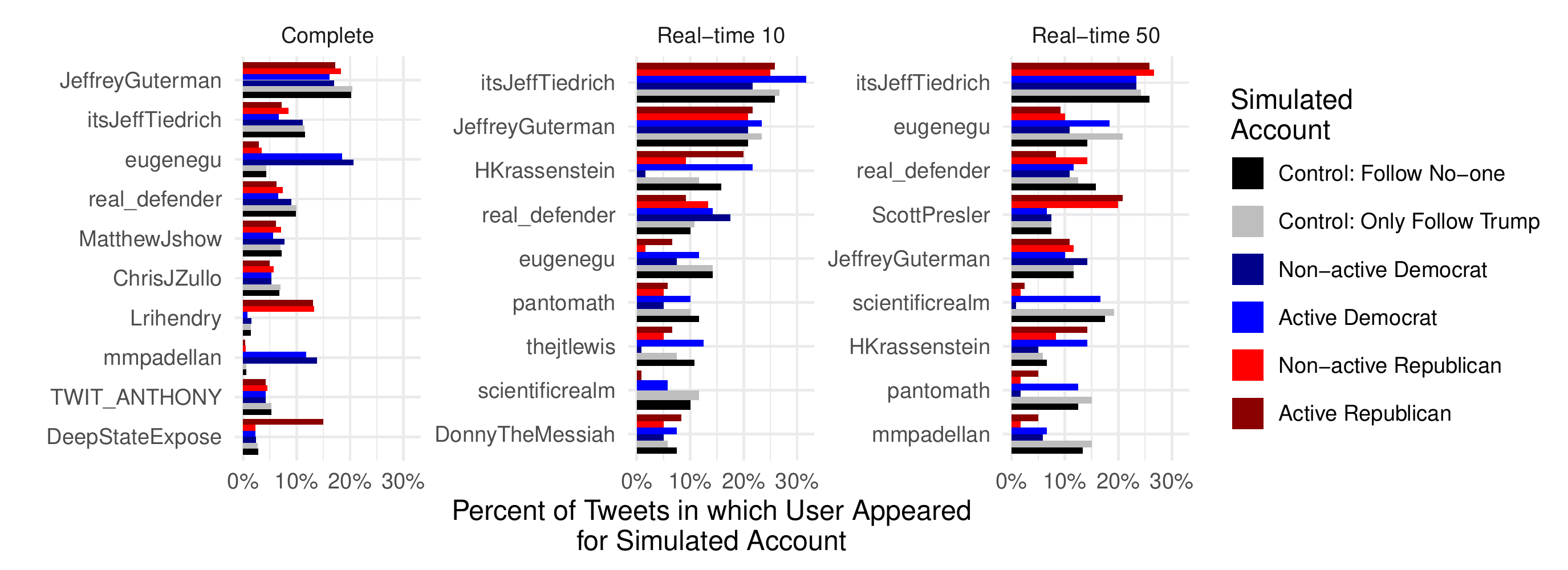}
    \caption{Each sub-plot, from left to right, presents the top 10 accounts (x-axis) in the given sample (subplot title) in terms of the average percentage of tweets in which they appear at least once across all simulated accounts. This average is then displayed as six different bars, one for each of the simulated accounts. The bars are ordered the same for each user, in the reverse order of the key. Note also that we remove from presentation one account with less than 5,000 followers to ensure privacy of highly personal accounts, and for presentation purposes, one account (\@LindaSuhler) seen at least once by only a subset of the simulated accounts.}
    \label{fig:u}
\end{figure*}

We find that individual Americans are well-represented in replies ``seen'' by the simulated accounts, and that while evidence exists of both source and content bias, it is limited in it's scope. Figure~\ref{fig:u} hints at these findings, which we expand upon in the subsections below. Figure~\ref{fig:u} shows, for the Complete sample and two Real-time samples, the top replier accounts in terms of their frequency of appearance.  For each top replier account, we also present six different bars, representing the percentage of tweets in which the replier appeared for each of the simulated accounts.

Of the top accounts appearing in Figure~\ref{fig:u}, none claim to be mass media accounts, and only one---\@MatthewJshow---claims to be a media outlet of any kind. Instead, accounts appearing most frequently in replies seen by the simulated accounts at least claim to be real 
Americans; for example, presenting in their profile descriptions as the ``mother of two awesome kids,'' and that ``My views are my own''.  However, such accounts could of course be only presenting as real people, and in reality be cyborg or even bot accounts. Thus, it is useful, as we do in the next section, to study whether panel member accounts, which we know to be associated with real people, also appear with relative frequency.

Similarly, Figure~\ref{fig:u} shows that some users, like \@itsJeffTiedrich, appear frequently and consistently for all six types of simulated accounts. In contrast, other accounts appear much more frequently for some accounts than others, and only in some time periods. For example, replies from \@ScottPresler are almost never shown to simulated Democrat or control accounts in the Complete or 50-minute Real-time sample, but appear with high frequency for simulated Republican accounts in these same samples.  In contrast, the account appears relatively frequently for all simulated accounts in the 10 minute Real-time sample.  These variations on specific accounts suggest a need to take a broader view of potential partisan biases in replies, which we do below by studying aggregate measures of source and content bias.



\subsubsection{Are individual Americans heard?}
\begin{figure}[t]
    \centering
    \includegraphics[width=.96\linewidth]{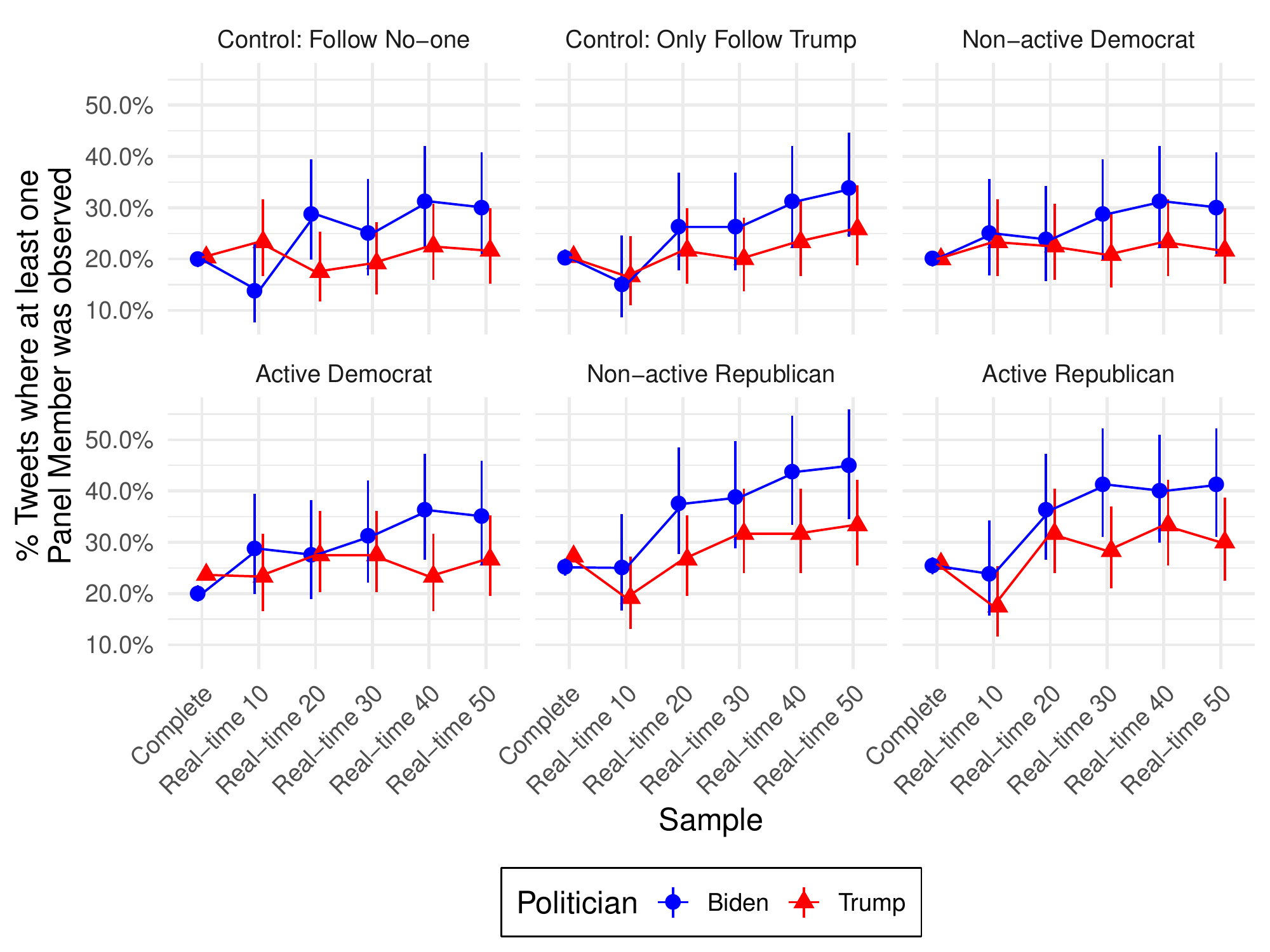}
    \caption{Across all tweets ``seen'' by the six different simulated accounts (different subplots, simulated account in top grey bar in label), the percentage (y-axis) that contain at least one reply from a panel member. The X-axis shows the six different samples - the five different Real-time samples (e.g. ``Real-time 10'' are replies collected 10 minutes after a tweet was sent) and the Complete sample.}
    \label{fig:heard}
\end{figure}

Averaging across samples and politicians, 22.7\% [22.4, 23.0] of the tweets seen by simulated accounts included a reply from at least one panel member. Some of this is due to the fact that some panel accounts are followed by the simulated accounts. After removing all tweets sent by panel accounts followed by the simulated users, however, we still find that 18.2\% [17.9,18.5] of viewed tweets contain a response from at least one panel member.  

Figure~\ref{fig:heard} displays the average proportion of Trump's tweets shown to each of the six different simulated accounts. To assess differences across sample, politician, and simulated account type, we carry out a logistic regression model with sampling period, Politician (Trump or Biden), and simulated account as the independent variables, and . Additionally, Figure~\ref{fig:heard} suggests an interaction between Politician and sampling period; differences between sampling periods appear to be much higher for Biden than for Trump. We thus include interaction terms for these effects in the model as well. Full results for the regression model can be found in Table~\ref{tab:reg_res} in the Appendix.  

We find that that Active Democrats, Active Republicans, and Non-active Republicans all see significantly more replies from panel members than the control accounts (p $<$ .0001 for all coefficients). Model results also suggest that a Trump tweet was approximately 6.5\% more likely than a Biden tweet to contain a reply from a panel member (p $<$ .001), and that there are minimal differences across time-period samples for Trump's tweets but significant differences for Biden.  Given the present day importance of Trump's tweets relative to Biden, however, we do not dive deeper into these differences across samples in the this work.

Results in this section thus cumulatively suggest that despite the panel representing a relatively small percentage of all real Americans on Twitter, panel members---and hence, real Americans--- are not entirely drowned out by organizations or political bots.  Moreover, at least for Trump, this finding is largely consistent across different time periods at which one might view a tweet, and across six different hypothesized types of Twitter users.  However, across samples, the panel is responsible for only 3.0\% of the total number of replies seen for Trump (and, coincidentally, 3.0\% for Biden as well). In the next section, we turn to an analysis of this full set of replies, and specifically, the source and content biases in them.

\subsubsection{Do Replies Create a ``Filter Bubble''?}

\begin{figure}[t]
    \centering
    \includegraphics[width=.98\linewidth]{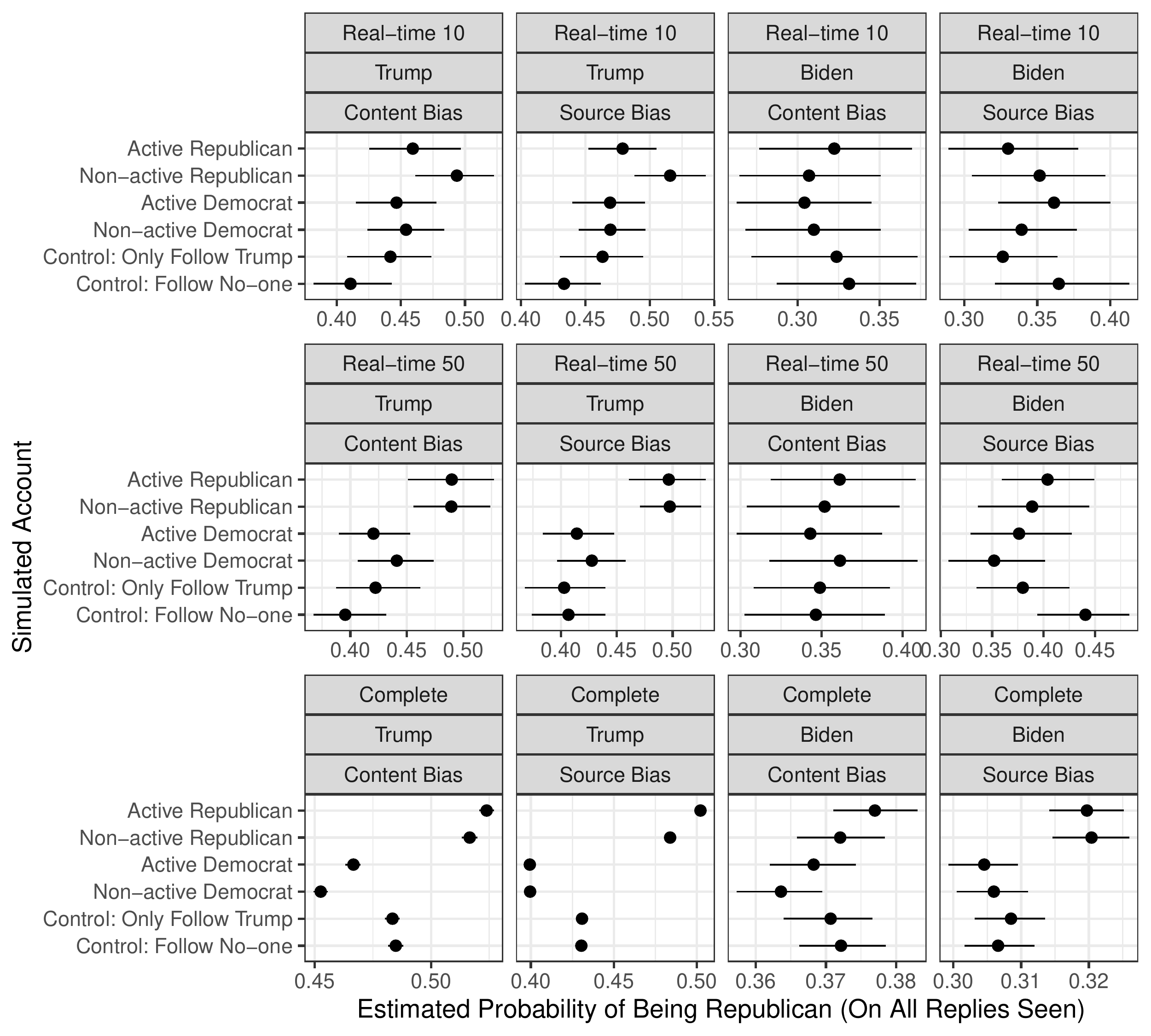}
    \caption{From left to right on a given row, the estimated source and content bias of replies to Trump and Biden, respectively. These measures are both given as the probability of a given source (reply text) being (sent by) a Republican.  From top to bottom, the three rows of plots  represent one of three different collection periods - the Real-time sample collected 10 minutes after tweets were sent, the Realtime sample collected 50 minutes after tweets were sent, and the Complete sample collecting replies for tweets from 2018 through 2020. }
    \label{fig:biasall}
\end{figure}

We find evidence of both source and content bias in replies seen by the simulated accounts. However, this bias is only \emph{statistically} significant 1) for replies to Trump's tweets (relative to Biden's), or 2) for both Trump and Biden, after a significant amount of time has passed since the tweet was sent.  Further, this effect is arguably only \emph{practically} significant for 3) Trump's tweets after a reasonable amount of time has passed since the tweet was sent, or 4) for Trump and Biden, for the relatively small subset of replies sent by accounts followed by the simulated users.

Figure~\ref{fig:biasall} presents evidence for claims 1), 2), and 3) above using a subset of the sampling periods. Results for all samples are presented in Figures~\ref{fig:sourcebias} (for source bias) and~\ref{fig:contentbias} (for content bias) in the Appendix. With respect to 1), Figure~\ref{fig:biasall} shows that statistically significant differences arise between our simulated Non-active Republican user and all other accounts only ten minutes after the reply is sent.  With respect to 2), in both the Real-time 50 minute sample and the Complete sample, statistically significant differences arise between the replies seen by simulated Democrat and Republican accounts for replies to both Trump and Biden. 

Perhaps most importantly, however, with respect to claim 3) above, only for Trump are these differences \emph{practically} significant. For example, Figure~\ref{fig:biasall} shows that our simulated Non-active Democrat is around 10\% more likely than the simulated Active Republican to see a reply to Trump that is sent by or contains content reflective of another Democrat. In contrast, even in the most extreme case for Biden, simulated Democrat accounts are only 1-2\% more likely than simulated Republican accounts to see replies in which the source or content also lean left politically.
Taken together, these findings suggest that Trump's tweets may uniquely provoke highly partisan responses that, in turn, are somewhat more likely t be shown to individuals who are likely to have a favorable view of the sender and content of that reply. This suggests that replies may indeed increase partisan framing of Trump's tweets. 

\begin{figure}[t]
    \centering
    \includegraphics[width=.98\linewidth]{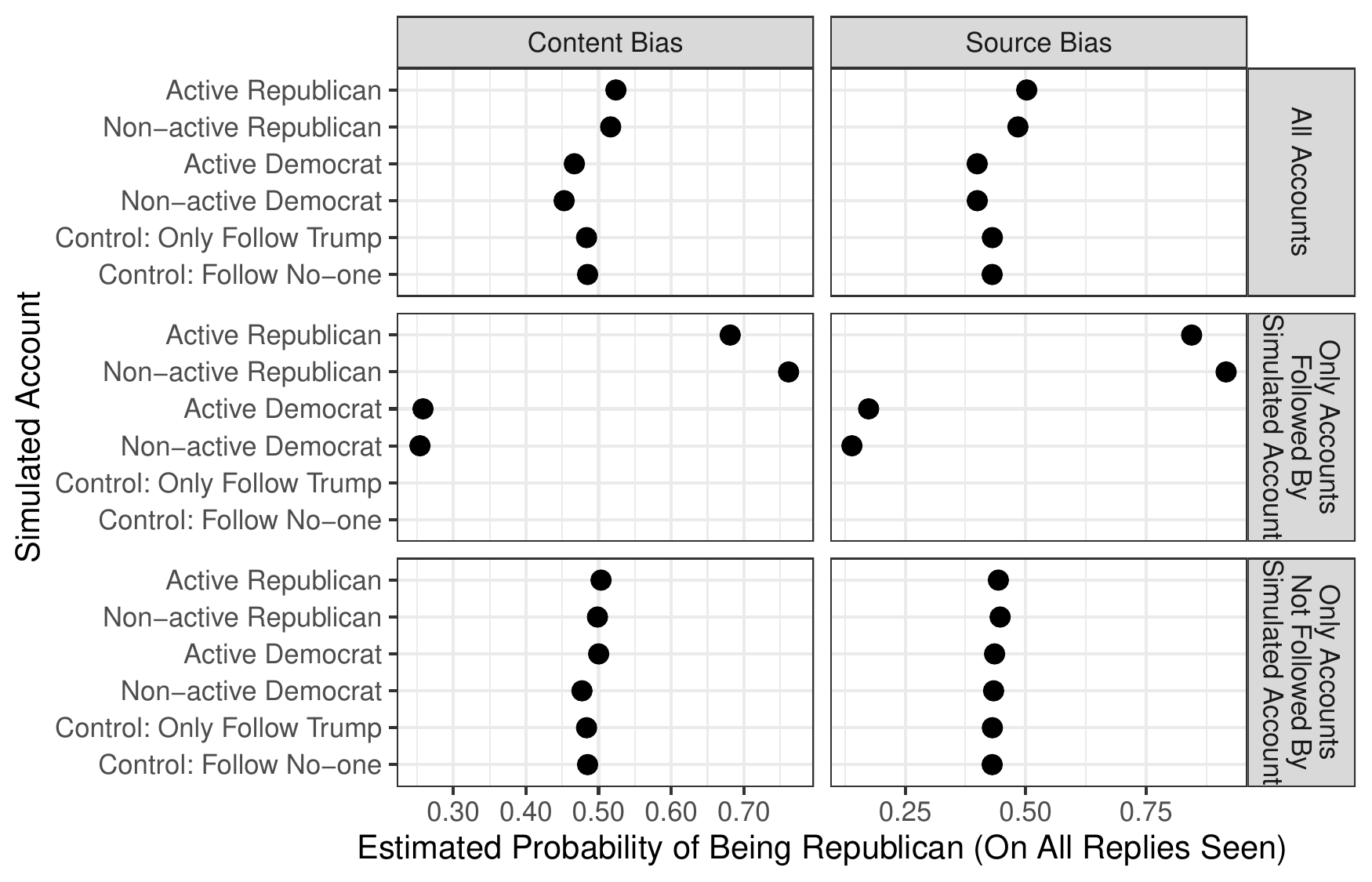}
    \caption{Content and Source bias (left and right columns, respectively) for the Complete Sample of replies to Trump's tweets for the six simulated users (x-axis). The different rows show different subsets of replies. The top row shows replies from all accounts, the middle from only accounts followed by the panel, and the bottom only from accounts not followed by the panel}
    \label{fig:biascomplete}
\end{figure}

However, Figure~\ref{fig:biascomplete} emphasizes that we should be cautious of attributing this effect directly to an ``algorithmic filter bubble''. The figure shows source and content bias for the most extreme case of bias- the Complete sample of Trump's tweets. We split these replies into three types - all replies (top row), only replies from accounts followed by the simulated user (middle row), and only replies from accounts \emph{not} followed by the simulated user.  Across the four non-control simulated accounts, between 9.0-10.1\% of replies seen were from accounts followed by the simulated user.

Figure~\ref{fig:biascomplete} shows drastic differences in the source and content biases of accounts followed by the simulated users. For replies to Trump, these differences are large enough to, in turn, generate the observed significant and practical differences in source and content bias across all replies seen in the top row of Figure~\ref{fig:biascomplete} and in the corresponding subplots of Figure~\ref{fig:biasall}. In contrast, while statistically significant differences in source and content bias arise between simulated Democrat and Republican users in replies from non-followed accounts, these differences are minimal, on the order of 2-3\%. Thus, even in the most extreme case of content and source bias, the practically significant difference in source and content bias across the political spectrum can be almost exclusively attributed to the appearance of replies from accounts followed by the viewing user.

These results align well conceptually with earlier findings from  \citeapos{robertson_auditing_2018}, who found that biases in Google search queries could be attributed to direct actions of the user (the query itself) rather than any black-box algorithmic behavior.  In our case, we do find, for Trump, statistically significant differences in replies shown to simulated users even when excluding accounts followed by those simulated users. Thus, there is evidence that Twitter users indirect signals to shape replies seen by accounts. However, as with prior work, the bulk of the observed source and content bias appears to come instead from the direct link between accounts followed and replies seen.

\section{Conclusion}

The present work presents the first attempt we are aware of to assess 1) who replies to President Trump's tweets, or 2) which replies are seen. We also show that findings for Trump's tweets are largely consistent with findings for tweets from Joseph Biden in the former case. In the latter case, we instead find some evidence that biases in replies shown to simulated accounts may exist only for Trump's tweets. This, we believe, can be reasonably attributed to the divisiveness that Trump has produced within American society, relative to Biden.  The methods we use to make these claims are straightforward, and our findings in the latter case are consistent across two different measures of bias. Finally, our results align with prior work in similar domains. All of this gives us further confidence in our findings.

However, our study does suffer from several notable limitations. First, we study replies to only two individuals. While President Trump and Joseph Biden are arguably, and currently, the most important individuals on Twitter in the U.S., leading to our decision, future work might explore replies to other public figures as well. Second, our analysis of individual Americans' replies is limited to a panel of users that we are able to link to voter registration records, and our analyses of demographics are based on self-reports in those records.  Finally, our simulated accounts are only representative of a certain subset of the population, and may not reflect results from the entirety of the broader public.

These limitations leave open several interesting and unanswered questions. For example, as noted, we use only following relationships as implicit signals of account partisanship. If Twitter's algorithm driving which replies are shown is driven by other forms of implicit signals, however, then algorithmically-curated partisanship in replies could still manifest.  Additionally, there are several questions we do not consider at all that may be of interest in future work, for example, how replies differ with respect to the content of Trump, Biden, or another politicians original tweet.

Nonetheless, our results have important ramifications for our understanding of how political content is consumed and responded to on Twitter. With respect to political voice and attention, we find that a small number of highly politically active individuals voice their opinion of and pay attention to a broad range of content produced by President Trump. However, the vast majority of individuals either choose not to express their voice through replies at all, or do so only to a very small subset of Trump and Biden's tweets. This is further evidence of why care must be taken in analyzing non-representative signals from social media \cite{hargittai_whose_2019}. 

With respect to algorithms and filter bubbles, we find that individuals that have similar partisan leanings may indeed see replies biased towards their political leaning. This seems to be especially true for tweets sent by Trump, relative to Biden, and after at least an hour or so has passed after the tweet was originally sent. However, we also find that these effects seems to be driven in large part directly by user decisions on whom to follow.   Like prior work, then, we find that future research may best be served studying why partisanship manifests in user decision-making online, rather than within recommendation algorithms.

\begin{small}
\bibliography{bib}
\bibliographystyle{aaai}
\end{small}

\section{Appendix}

\begin{figure}[th]
    \centering
    \includegraphics[width=.98\linewidth]{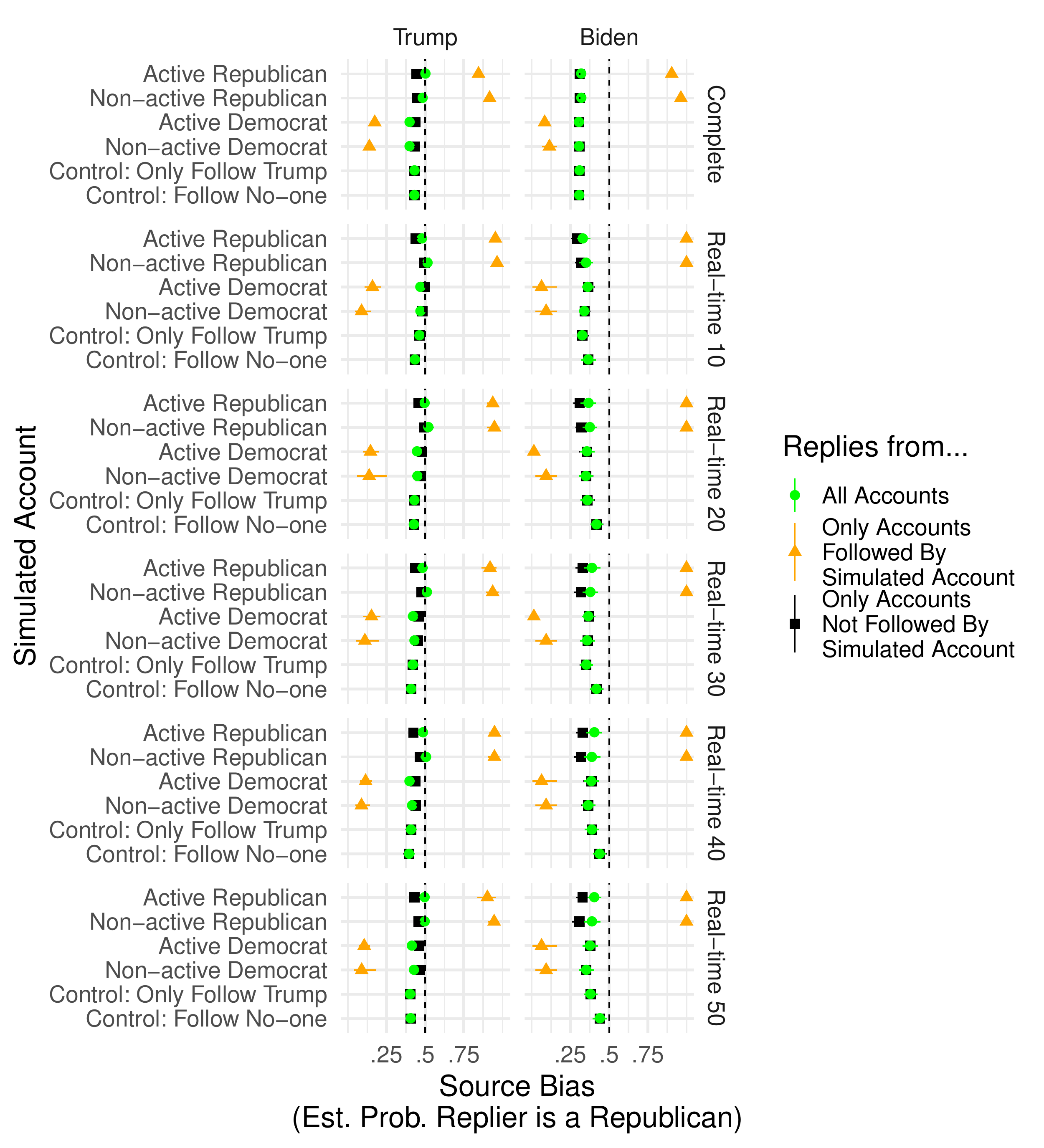}
    \caption{Mean estimated source bias across all tweets (x-axis)  seen by each simulated account (y-axis) for Biden and Trump (separate columns) during each collection period (separate rows). Confidence intervals are 95\% bootstrapped estimates. Results are shown for all replies (green), only accounts not followed by any simulated account (black), and only accounts that \emph{are} followed by a simulated account (orange).}
    \label{fig:sourcebias}
\end{figure}

\begin{figure}[t]
    \centering
    \includegraphics[width=.98\linewidth]{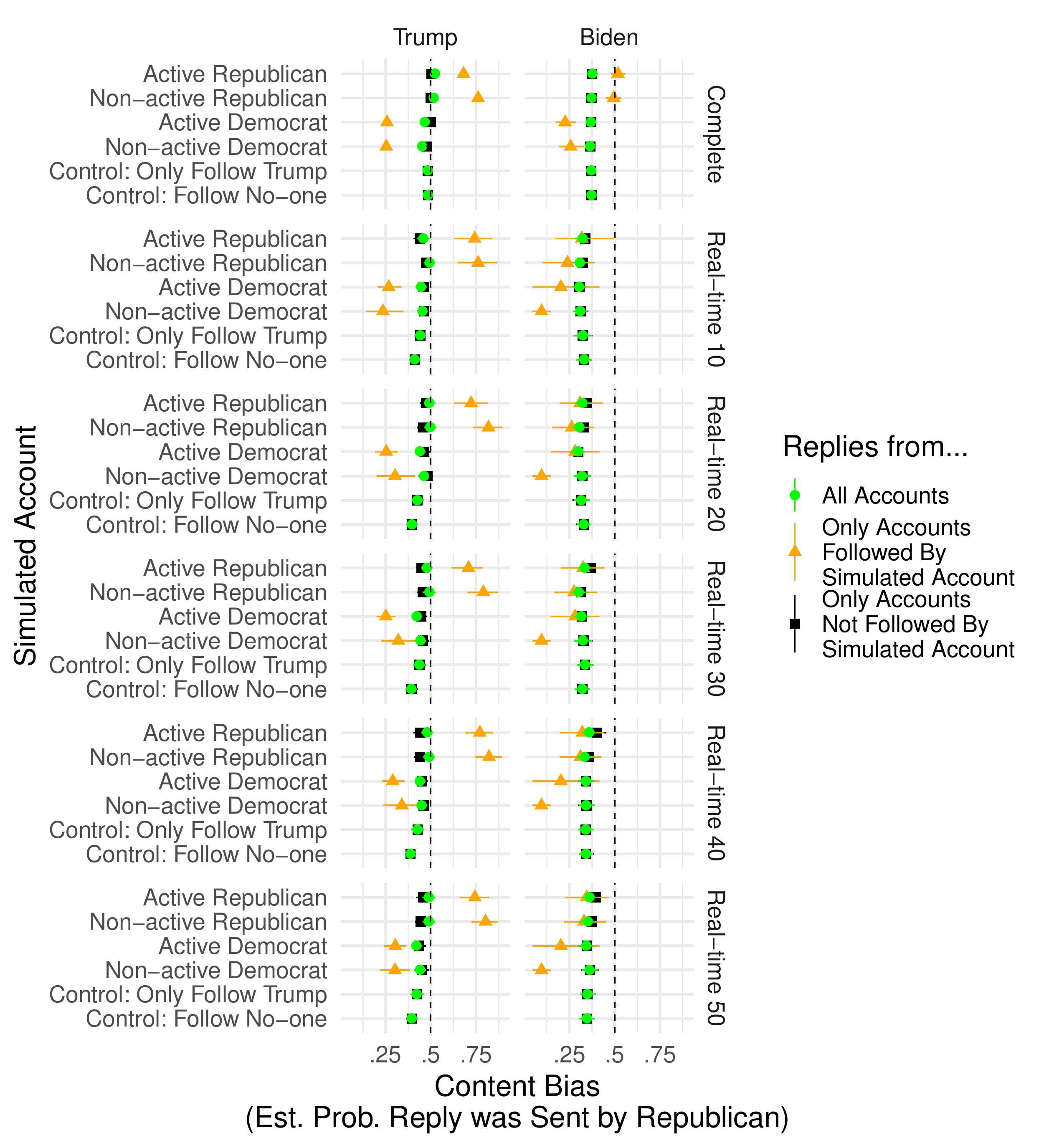}
    \caption{Mean estimated content bias across all tweets (x-axis)  seen by each simulated account (y-axis) for Biden and Trump (separate columns) during each collection period (separate rows). Confidence intervals are 95\% bootstrapped estimates. Results are shown for all replies (green), only accounts not followed by that simulated account (black), and only accounts that \emph{are} followed by that simulated account (orange).}
    \label{fig:contentbias}
\end{figure}

\begin{table}[t] \centering 
\small
\begin{tabular}{p{5.2cm}p{1.5cm}} 
\\[-1.8ex]\hline 
\\[-1.8ex] & \emph{Coefficient} \\ 
\hline \\[-1.8ex] 
(Sim. Account) Control: Only Follow Trump & 0.015 \\ 
  & (0.031) \\ 
  & \\ 
(Sim. Account) Non-active Democrat & $-$0.001 \\ 
  & (0.031) \\ 
  & \\ 
(Sim. Account) Active Democrat & 0.169$^{***}$ \\ 
  & (0.031) \\ 
  & \\ 
(Sim. Account) Non-active Republican & 0.379$^{***}$ \\ 
  & (0.030) \\ 
  & \\ 
(Sim. Account) Active Republican & 0.326$^{***}$ \\ 
  & (0.030) \\ 
  & \\ 
(Sent by) Trump & 0.065$^{***}$ \\ 
  & (0.022) \\ 
  & \\ 
(Sample) Real-time 10 & 0.006 \\ 
  & (0.112) \\ 
  & \\ 
(Sample) Real-time 20 & 0.434$^{***}$ \\ 
  & (0.102) \\ 
  & \\ 
(Sample) Real-time 30 & 0.522$^{***}$ \\ 
  & (0.100) \\ 
  & \\ 
(Sample) Real-time 40 & 0.691$^{***}$ \\ 
  & (0.098) \\ 
  & \\ 
(Sample) Real-time 50 & 0.700$^{***}$ \\ 
  & (0.097) \\ 
  & \\ 
(Sent by)Trump x (Sample)Real-time 10 & $-$0.144 \\ 
  & (0.146) \\ 
  & \\ 
(Sent by)Trump x (Sample)Real-time 20 & $-$0.340$^{**}$ \\ 
  & (0.134) \\ 
  & \\ 
(Sent by)Trump x (Sample)Real-time 30 & $-$0.428$^{***}$ \\ 
  & (0.133) \\ 
  & \\ 
(Sent by)Trump x (Sample)Real-time 40 & $-$0.509$^{***}$ \\ 
  & (0.130) \\ 
  & \\ 
(Sent by)Trump x (Sample)Real-time 50 & $-$0.503$^{***}$ \\ 
  & (0.129) \\ 
  & \\ 
 Constant & $-$1.434$^{***}$ \\ 
  & (0.028) \\ 
  & \\ 
\hline \\[-1.8ex] 
Observations & 76,344 \\ 
Log Likelihood & $-$40,569.140 \\ 
Akaike Inf. Crit. & 81,176.280 \\ 
\hline 
\hline \\[-1.8ex] 
\end{tabular} 
  \caption{Results of a logistic regression model to predict whether or not at least one reply from a panel member would be observed in a given politician's tweet seen in a given sample by a given simulated account.\\
  $^{*}$p$<$0.1; $^{**}$p$<$0.05; $^{***}$p$<$0.01
  }
  \label{tab:reg_res} 
\end{table} 

Table~\ref{tab:reg_res} displays coefficients for the logistic regression model described in the main text to predict whether or not the replies seen for a given tweet from Trump or Biden would show at least one reply from a panel member. The reference simulated account in the model is the Control account following no one, the reference sample is the Complete sample, and the reference politician is Biden.  Note that the interaction terms in the regression model essentially cancel out the main effects of the sample seen for Biden. This implies, as noted in the main text, that across samples, replies to Trump's tweets were roughly equally likely to contain a reply from a panel member.

\end{document}